\documentclass[twocolumn,prl,aps,tightenlines,superscriptaddress,a4paper]{revtex4}

\usepackage{amsmath}
\usepackage{amssymb}
\usepackage{graphicx}

\newcommand{\bra}[1]{\mbox{$\langle #1|$}}
\newcommand{\ket}[1]{\mbox{$|#1\rangle$}}

\newcommand{\ketbra}[2]{\mbox{$|#1\rangle\langle #2|$}}

\begin{document}
\title{Information complementarity in quantum physics}

\author{A. Fedrizzi}
\affiliation{Department of Physics and Centre for Quantum Computer Technology, University of Queensland, Brisbane 4072, Australia}
\author{B. \v{S}kerlak}
\affiliation{Department of Physics and Centre for Quantum Computer Technology, University of Queensland, Brisbane 4072, Australia}
\author{T. Paterek}
\affiliation{Centre for Quantum Technologies and Department of Physics, National University of Singapore, 3 Science Drive 2, 117542 Singapore}
\author{M. P. de Almeida}
\affiliation{Department of Physics and Centre for Quantum Computer Technology, University of Queensland, Brisbane 4072, Australia}
\author{A. G. White}
\affiliation{Department of Physics and Centre for Quantum Computer Technology, University of Queensland, Brisbane 4072, Australia}

\begin{abstract}
We demonstrate that the concept of information offers a more complete description of complementarity than the traditional approach based on observables. We present the first experimental test of information complementarity for two-qubit pure states, achieving close agreement with theory; We also explore the distribution of information in a comprehensive range of mixed states. Our results highlight the strange and subtle properties of even the simplest quantum systems: for example, entanglement can be increased by reducing correlations between two subsystems.
\end{abstract}

\maketitle
Complementarity reveals trade-offs between knowledge of physical observables. The best known example is wave-particle duality: a single quantum system may exhibit wave and/or particle properties, depending on the experimental context. For a system in a two-mode interferometer, this is quantitatively expressed by the fact that the interference visibility $\mathcal{V}$ and the mode predictability $\mathcal{P}$ have to satisfy \cite{greenberger1988swa,englert1996fva}:
\begin{equation}
\mathcal{V}^{2}+\mathcal{P}^{2}\leq1.
\label{eq:duality}
\end{equation}
High quality interference comes at the expense of the impossibility of predicting the path of the system, and vice versa, a phenomenon which has been demonstrated in a host of physical systems \cite{schwindt1999qwp,chang2008qmc,zeilinger1988sad,durr1998fvw,gerlich2007kdt,kolesov2009wpd}. Relation (\ref{eq:duality}) equals unity only for pure, single-particle, quantum states. For mixed states, the left hand side is always less than one and it can even reach zero \cite{schwindt1999qwp}, which means that there is no knowledge about whether the system behaves as a particle or a wave. 

This \emph{knowledge deficit} for a single particle in a mixed state can be attributed to entanglement with a second particle. Instead of limiting ourselves to the \emph{local} observables $\mathcal{V}$ and $\mathcal{P}$, one includes the \emph{non-local} observable $\mathcal{C}$---the concurrence \cite{hill1997epq}, a measure for entanglement, and finds~\cite{jakob2003qcr,demelo2007qnc}:
\begin{equation}
(\mathcal{V}^{2}+\mathcal{P}^{2})_{\mathrm{local}}+(\mathcal{C}^{2})_{\mathrm{corr}}\leq1. 
\label{eq:triality}
\end{equation}
\begin{table}[!bp]
\vspace{-4mm}
\caption{Three families of quantum states which we use to explore information complementarity. i) Pure states $\rho_{\textsc{pure}}$; ii) highly-entangled mixed-states $\rho_{\textsc{werner}}$ and $\rho_{\textsc{mems}}$; and iii) separable states $\rho_{\textsc{as}}$, $\rho_{\textsc{s}}$}
\vspace{-2mm}
\begin{center}
\begin{tabular}{@{\vrule height 10.5pt depth4pt  width0pt}l l l@{~~~} l}
i) & $\rho_{\textsc{pure}}{=}$ & $\ketbra{\psi}{\psi}$; $\ket{\psi} = \cos \alpha \ket{00}+\sin \alpha \ket{11}$ & $\alpha\in[0,\pi/4]$ \\
\hline
ii) & $\rho_{\textsc{werner}}{=}$ & $p\ket{\psi^-}\bra{\psi^-} + \frac{1-p}{4}{\openone}_{4}$ & $p\in[0,1]$ \\
& $\rho_{\textsc{mems}}{=}$ & $p\ket{\phi^+}\bra{\phi^+}+(1-p)\ket{10}\bra{10}$ & $ p\in[\frac{2}{3},1]$ \\
& & $0\leq S(\rho)\lesssim0.92$ \\
\hline
iii) & $\rho_{\textsc{as}}^{(1)}{=}$ & $(p\ket{0}\bra{0}+\frac{1-p}{2}{\openone}_{2})\otimes\ket{0}\bra{0}$ & $p\in[0,1]$ \\
& & $0\leq S(\rho)\leq1$ \\
& $\rho_{\textsc{as}}^{(2)}{=}$ & ${\openone}_{2}\otimes(q\ket{0}\bra{0}+\frac{1-q}{2}{\openone}_{2})$ & $q\in[0,1]$ \\
&  & $1\leq S(\rho)\leq2$ \\
& $\rho_{\textsc{s}}{=}$ &$p\ket{00}\bra{00}+\frac{1-p}{4}{\openone}_{4}$ & $p\in[0,1]$\\
\end{tabular}
\end{center}
\label{tab:states}
\vspace{-4mm}
\end{table}
The complementarity is now between the \emph{local} properties of individual subsystem and its \emph{correlations} with another subsystem. For pure two-particle states, relation (\ref{eq:triality}) has been experimentally tested to some extent in \cite{abouraddy2001dco,peng2005qcm}---albeit in its early form of interferometric complementarity \cite{jaeger1993cop}. In \cite{salles2008eid}, Eq. \ref{eq:triality} has been measured for a `system' qubit coupled to an `evironment' qubit via an amplitude damping channel. Once again, for the case of mixed states, (\ref{eq:triality}) does not saturate its bound. One could of course explain this by entanglement to yet another particle and so on, ad infinitum \cite{HH}. It would however be preferable to precisely identify the quantities involved in complementarity without resorting to virtual higher-dimensional Hilbert spaces.

In this Letter we experimentally study a version of complementarity between two quantum bits which does not require infinite regression. It is phrased in terms of information and is symmetric and exact, i.e. all subsystems are treated on equal footing and the relations are saturated also for mixed states. We experimentally test the complementarity for pure two-qubit states and explore distribution of information in many interesting mixed two-qubit states, see table \ref{tab:states}. Surprisingly at first glance, this shows that more entanglement can occur with less correlations and allows understanding why traditional complementarity relations are not exact.
\begin{figure}[h!tbp]
\begin{center}
\includegraphics[width=0.38\textwidth]{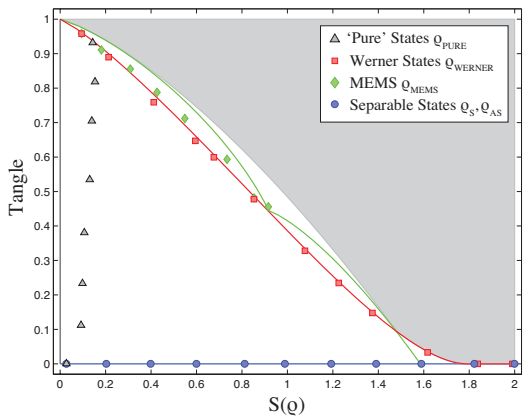}
\caption{Tangle-von-Neumann-entropy plane, showing the 48 experimentally created two-photon states. Error bars are smaller than the symbol size. The lines are predictions for the ideal states, Table \ref{tab:states}. Data points for $\rho_{\textsc{as}}$ and $\rho_{\textsc{s}}$ overlap. The shaded area represents unphysical states, states on the boundary are rank-3 MEMS \cite{wei2003mev}. \vspace{-8mm}}
\label{fig:tangent-plane}
\end{center}
\end{figure}

Information complementarity goes back to the insight that information is physical \cite{landauer1996pni} and that the amount of information in a two-level quantum system (or \emph{qubit}) is limited to $1$ bit~\cite{zeilinger1999fpq,brukner1999oii}. Complementarity may now arise because there is insufficient information to simultaneously specify the results of different measurements that can be performed upon a quantum system. The single-particle complementarity relation (\ref{eq:duality}) reflects this by being bounded to $1$ (bit), a value which is saturated for a \emph{pure} qubit. For a mixed state, the left hand side of (\ref{eq:duality}) is equal to a suitable measure of the reduced information content of the system \cite{brukner1999oii}.

This provides an alternative interpretation of the two-particle complementarity, Eq.~\ref{eq:triality}. The total information content of two particles, $I_{\mathrm{total}}$, can be split into information stored in subsystems, $I_{\mathrm{local}}$, and a remaining part which we call the \emph{correlation information}, $I_{\mathrm{corr}}$, giving rise to the following relation \cite{oppenheim2003mea}:
\begin{equation}
I_{\mathrm{local}}+I_{\mathrm{corr}}=I_{\mathrm{total}}.
\label{eq:complement-information}
\end{equation}
The natural way to quantify the information of a given quantum state is to use its entropy. The information measure originally proposed in \cite{brukner1999oii} is linked to the so called \emph{linear entropy}, and, as we show in the supplementary online material \cite{SOM}, for pure states~(\ref{eq:complement-information}) is in this case equivalent to (\ref{eq:triality}). Unfortunately, for certain mixed two-particle states, the linear entropy measure leads to a \emph{negative} $I_{\mathrm{corr}}$ \cite{SOM}. We therefore adopt the \emph{von Neumann} information $I(\rho){=}\log d{-}S(\rho)$, where $S(\rho){=}{-}\mathrm{Tr}(\rho\log_{2}\rho)$ is the von Neumann entropy of a $d$-dimensional system described by the density matrix $\rho$. The correlation information is then the \emph{quantum mutual information}, $I_{\mathrm{corr}}{=}I_{\mathrm{total}}{-}I_{\mathrm{local}}{=}S(\rho_{a}){+}S(\rho_{b}){-}S(\rho_{ab})$, which is non-negative for all physical states and a measure for the total correlations present in a quantum state \cite{groisman2005qct}. 

Unlike traditional wave-particle duality, the information approach (\ref{eq:complement-information}) has not been explored in any experimental system to date. We now investigate information complementarity for a range of two-photon quantum states. For pure states, we can test complementarity relation (\ref{eq:complement-information}); for mixed states, it will allow us to highlight the different types of correlations present in these systems. Our states can be grouped into three families, Table I: i) ``pure'' states, i.e. our best experimental approximation to pure; ii) two classes of highly-entangled mixed states---Werner and maximally-entangled mixed-states \cite{munro2001met,wei2003mev} (MEMS); and iii) two classes of separable states. These states were chosen because they represent (or are close to) the boundaries of the physical parameter space for two-qubit states in the tangle-entropy plane, Fig.~\ref{fig:tangent-plane}.
\begin{figure}[!t]
\begin{center}
\includegraphics[width=0.42\textwidth]{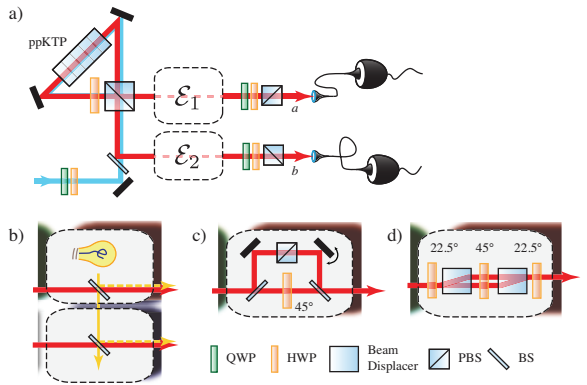}
\caption{Experimental scheme. a) A source creates pure two-photon states with a tunable degree of entanglement. Channels $\mathcal{E}_{1}$ and $\mathcal{E}_{2}$ introduce mixing before state tomography is performed. b) For the creation of $\rho_{\textsc{werner}}$ and $\rho_{\textsc{s}}$, Table \ref{tab:states}, an incandescent light source with tunable intensity is reflected into the setup. The increase in accidental coincidence detections resembles white noise. c) For MEMS state creation \cite{aiello2007mem}, a beamsplitter (BS) is introduced. The transmitted beam polarisation is rotated by $90\symbol{23}$ and the reflected beam is polarized at a polarizing beam splitter (PBS) before the beams are recombined incoherently. Two steering mirrors control the splitting ratio between the two beams and thus parameter $p$ in $\rho_{\textsc{mems}}$, Table \ref{tab:states}. This technique allows the creation of $\rho_{\textsc{mems}}$ in a range of $0{\leq}S(\rho){\lesssim}0.92$. The remaining states can be covered by dephasing the second photon once $p{=}2/3$ is reached. In practice, the initial tangle in our experiment was too low to create states with significantly higher entanglement than the corresponding $\rho_{\textsc{werner}}$. d) Dephasing channel for the creation of $\rho_{\textsc{as}}$. Jamin-Lebedeff interferometers \cite{obrien2003dao} introduce optical path delays between orthogonal polarization components of incoming photons in a given basis. Two interferometers are used to first individually decohere photon $1$ from zero to fully mixed and then photon $2$. \vspace{-4mm} }
\label{fig:scheme-combined}
\end{center}
\end{figure}
The experimental scheme is depicted in Fig.~\ref{fig:scheme-combined}. A source based on a polarization Sagnac loop \cite{kim2006pss,fedrizzi2007wtf} produces two-photon states close to the ideal form $\ket{\psi_{ab}}{=}\cos \alpha \ket{H_{a}H_{b}}+\sin \alpha \ket{V_{a}V_{b}}$, where the logical qubit states `0' and `1' are encoded into horizontal (H) and vertical (V) polarizations of the photons $a$ and $b$. The degree of entanglement is determined by $\alpha$, which is set by the pump laser polarization \cite{kim2006pss}. The resulting photons pass through two channels, $\mathcal{E}_{1}$ and $\mathcal{E}_{2}$, Fig. \ref{fig:scheme-combined}. We perform both full single-qubit state tomography \cite{james2001mq} on the individual photons $a$ and $b$---from which we reconstruct $\rho_{a}$ and $\rho_{b}$---and, separately, two-qubit tomography \cite{james2001mq} on the two-photon state, which yields $\rho_{ab}$. From these density matrices, we can readily compute properties such as the tangle $\mathcal{T}$---the concurrence $\mathcal{C}$ squared---or entropy. The $48$ created states are shown in Fig. \ref{fig:tangent-plane}: For each, we calculate $I_{\mathrm{a}}{=}I(\rho_{a})$ and $I_{\mathrm{b}}{=}I(\rho_{b})$ to obtain $I_{\mathrm{local}}{=}I_{\mathrm{a}}{+}I_{\mathrm{b}}$ and $I_{\mathrm{total}}{=}2{-}S(\rho_{ab})$ respectively. For the correlation information $I_{\mathrm{corr}}$, we subtract $I_{\mathrm{local}}$ from $I_{\mathrm{total}}$.
\begin{figure}[!t]
\begin{center}
\includegraphics[width=0.36\textwidth]{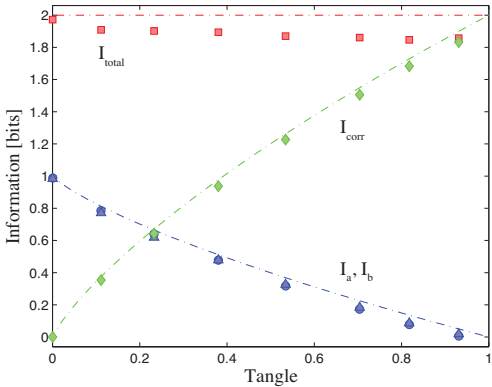} 
\caption{Information complementarity for ``pure'' states $\rho_{\textsc{pure}}$. Ideally, Eq. \ref{eq:complement-information} is in this case $I_{\textrm{a}}{+}I_{\textrm{b}}{+}I_{\textrm{corr}}{=}2$. The involved quantitites are measured independently, our result therefore represents a genuine test of complementarity. Ideal pure states are represented by lines. Our data ($\bullet{=}I_{\textrm{a}}$, $\blacktriangle{=}I_{\textrm{b}}$, $I_{\textrm{corr}}{=}\blacklozenge$, $I_{\textrm{total}}{=}\blacksquare$) falls short of these due to experimental imperfections. Error bars are smaller than symbol size. \vspace{-4mm}}
\label{fig:pure2}
\end{center}
\end{figure}

Information complementarity for our experimental states $\rho_{\textsc{pure}}$, is shown in Fig. \ref{fig:pure2}. When $\mathcal{T}$ vanishes, the total information is stored exclusively in the individual subsystems. As the entanglement increases, so does $I_{\mathrm{corr}}$ and as $\mathcal{T}$ approaches $1$, $I_{\mathrm{local}}$ goes to zero. This phenomenon is strictly quantum---a \emph{pure} \emph{classical} state cannot contain any mutual information. More importantly, for pure quantum states, $I_{\mathrm{corr}}$ corresponds to the \emph{entanglement of formation} \cite{horodecki2003lir}. This means that information complementarity reduces to a sum of entanglement and local information, each of which can be measured independently, and thus (\ref{eq:complement-information}) has the same form as (\ref{eq:triality}), the only difference being that (\ref{eq:complement-information}) is symmetric with respect to the subsystems, i.e. it accounts for the local properties of both subsystems $a$ and $b$. Because any pure entangled two-qubit state can be obtained from the Schmidt form $\rho_{\mathrm{pure}}$ via local operations, the data in Fig.~\ref{fig:pure2} represents a conclusive test of (\ref{eq:complement-information}) and (\ref{eq:triality}) for \emph{all} pure two-qubit states (see also \cite{SOM}).

\begin{figure}[htbp]
\begin{center}
\includegraphics[width=0.36\textwidth]{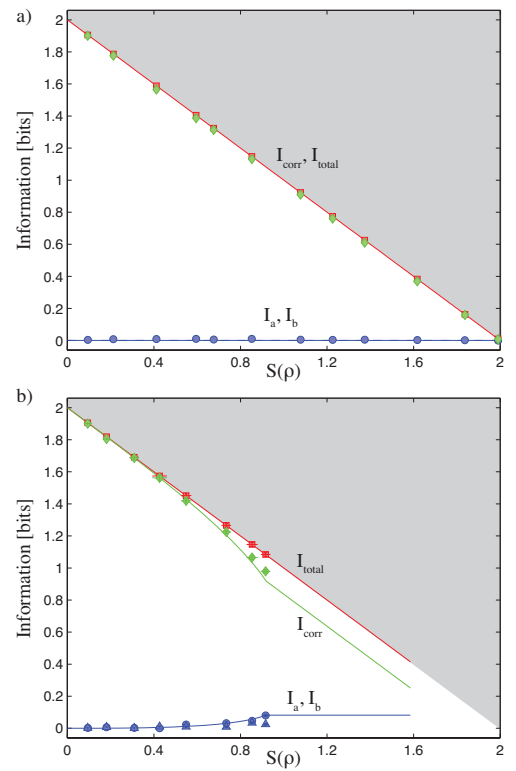} 
\caption{Information content in a) Werner states $\rho_{\textsc{werner}}$ and b) maximally-entangled mixed-states $\rho_{\textsc{mems}}$. a) The entire information content of $\rho_{\textsc{werner}}$ is stored in correlations. b) Even though $\rho_{\textsc{mems}}$ have more entanglement (cf. Fig.~\ref{fig:tangent-plane}) than $\rho_{\textsc{werner}}$, they have less $I_{\mathrm{corr}}$ ($\blacklozenge$). Solid lines represent ideal states. Error bars for a) are smaller than the symbol size ($\blacksquare{=}I_{\mathrm{total}}$, $\blacktriangle{=}I_{a}$, $\bullet{=}I_{b}$). For b), they were obtained assuming Poissonian count statistics. \vspace{-6mm}}
\label{fig:werms}
\end{center}
\end{figure}

At this stage it is tempting to identify $I_{\mathrm{corr}}$ with entanglement in general. We show experimentally that this is not the case by examining highly entangled mixed states. Werner states, $\rho_{\textsc{werner}}$, are a statistical mixture of a maximally entangled state and white noise. The individual qubits in this state are always fully mixed, $I_{\mathrm{local}}{=}0$, and their entire information content is stored in $I_{\mathrm{corr}}$, as one can see in Fig.~\ref{fig:werms}a). Werner states are separable for a high noise admixture ($\mathcal{T}{=}0$ for $S(\rho_{\textsc{werner}}){>}1.8$, Fig.~\ref{fig:tangent-plane}) but their $I_{\mathrm{corr}}$ does not vanish correspondingly, emphasizing that it does not measure entanglement in this case. 

By comparing Werner to MEMS states, it is straightforward to show that $I_{\mathrm{corr}}$ is not even a monotonic function of entanglement. Out of the several different existing classes \cite{wei2003mev} of MEMS, we chose to create rank-2 MEMS, $\rho_{\textsc{mems}}$, using the method from \cite{aiello2007mem}, illustrated in Fig.~\ref{fig:scheme-combined}b). As one can see in Fig.~\ref{fig:werms}b), the local information contents of these states are non-zero. For any value of $S$, even though they are more entangled (cf. Fig.~\ref{fig:tangent-plane}), $I_{\mathrm{corr}}$ for $\rho_{\textsc{mems}}$ is \emph{lower} than for the corresponding $\rho_{\textsc{werner}}$. In particular, we find that for $I_{\mathrm{total}}{=}1.4$ bits the Werner state has more mutual information $I_{\mathrm{corr}}{=}1.387{\pm}0.001$, and less tangle $\mathcal{T}{=}0.647{\pm}0.004$, than the corresponding MEMS, where $I_{\mathrm{corr}}{=}1.315{\pm}0.018$ and $\mathcal{T}{=}0.667{\pm}0.007$. Strikingly, the higher entanglement of MEMS states compared to Werner states coincides with a relative \emph{decrease of correlations}. In our experiment, this difference is small, due to the difficulties in producing high quality MEMS. There are however states for which the effect is far more pronounced \cite{SOM}.
\begin{figure}[!tbp]
\begin{center}
\includegraphics[width=0.36\textwidth]{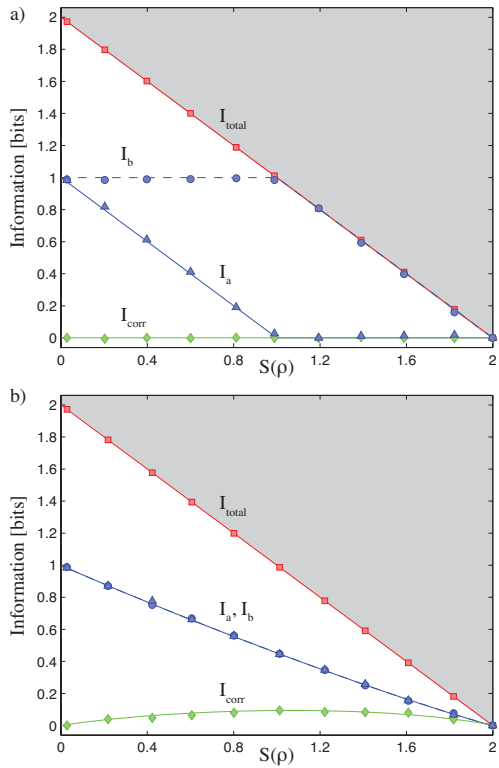} 
\caption{Information content in a) asymmetrically, $\rho_{\textsc{as}}$ and b) symmetrically mixed product states, $\rho_{\textsc{s}}$. Both states are separable but $\rho_{\textsc{s}}$ has non-zero correlations $I_{\textrm{corr}}$. Lines represent the information content of the ideal states, data points are measured values ($\blacksquare{=}I_{\mathrm{total}}$, $\blacklozenge{=}I_{\mathrm{corr}}$, $\blacktriangle{=}I_{a}$, $\bullet{=}I_{b}$). All error bars are smaller than symbol size. \vspace{-6mm}}
\label{fig:prod2}
\end{center}
\end{figure}

Our third example shows that aspects of complementarity between local information and correlations are already present in classical states, i.e. states which are mixtures of locally distinguishable states. We consider two classes of mixed separable states $\rho_{\textsc{as}}$ and $\rho_{\textsc{s}}$, Table \ref{tab:states}. The first class $\rho_{\textsc{as}}$ represents two individually dephased photons. The second class $\rho_{\textsc{s}}$ consists of an initially pure product state with an admixture of white noise, see Fig.~\ref{fig:scheme-combined}d) for experimental details. 

In Fig.~\ref{fig:prod2}a) one can clearly see that product states with individually added noise, $\rho_{\textsc{as}}$ do not contain any correlations, $I_{\mathrm{corr}}{=}0$. In contrast, states with globally added white noise, $\rho_{\textsc{s}}$, contain $I_{\mathrm{corr}}{=}0.094{\pm}0.023$ bits of mutual information at $I_{\mathrm{total}}{=}1$ bit, while $I_{\mathrm{local}}$ is reduced accordingly, Fig.~\ref{fig:prod2}b).

In conclusion, we performed a test of information complementarity and traditional complementarity for pure two-qubit states and showed that the two approaches can be reconciled in this case. For mixed states, we measured how information is distributed in a quantum system which allowed us to conclude that correlations are not a monotonic function of entanglement. This suggests what is missing in traditional complementarity relations like (\ref{eq:triality}). They are not exact because they do not take into account other correlations than those due to entanglement. It remains an open question if a traditional complementarity relation---based on directly observable quantities---can be formulated, which also includes classical correlations and quantum discord \cite{ollivier2001qdm, henderson2001cqt} or dissonance \cite{modi2009req}.

\begin{acknowledgments}
We thank R \L{}apkiewicz, BP Lanyon, {\v C} Brukner and A Zeilinger for valuable input. This work was supported by the ARC Discovery \& Fed.~Fellow programs and an IARPA-funded U.S. Army Research Office contract. TP acknowledges support from the National Research Foundation \& Ministry of Education, Singapore.
\end{acknowledgments}


\newpage

\section{Supplementary material: Information complementarity in quantum physics}

\setcounter{figure}{5}
\setcounter{equation}{3}

\subsection{Complementarity from linear entropy}
\begin{figure}[!b]
\begin{center}
\includegraphics[width=0.34\textwidth]{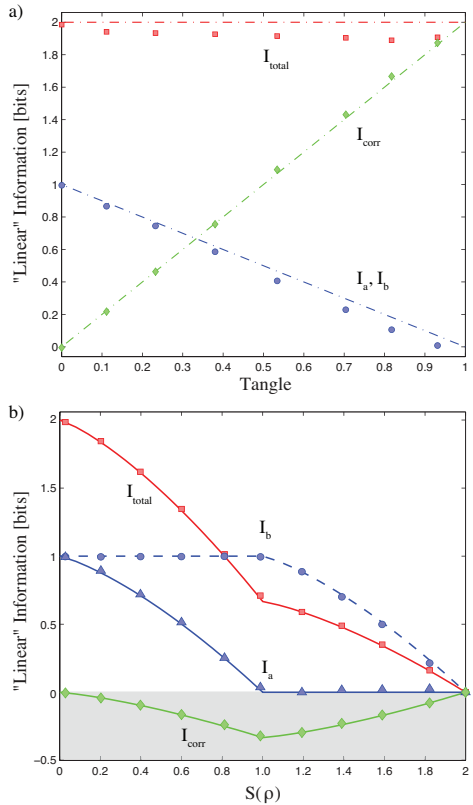}
\caption{Experimental verification of traditional complementarity relation (2) of the main text. a) For $\rho_{\textsc{pure}}$, there is a linear relation between information in correlations $I_{\mathrm{corr}}$ ($\blacklozenge$) and tangle. b) For the mixed state $\rho_{\textsc{as}}$, $I_{\mathrm{corr}}$  assumes negative values (shaded area) for the full mixing range.  ($I_{\mathrm{total}}{=}\blacksquare, I_{\mathrm{a}}{=}\blacktriangle, I_{\mathrm{b}}{=}\bullet$, error bars smaller than symbols).}
\label{fig:linent}
\end{center}
\end{figure}

Here we show that for pure states the complementarity relation (2) of the main text is a special case of the information complementarity
\begin{equation}
I_{\mathrm{local}} + I_{\mathrm{corr}} = I_{\mathrm{total}}.
\label{SOM_CI}
\end{equation}
For the moment, we quantify information using the measure of \cite{si:brukner1999oii}, which is based on linear entropy \cite{si:wei2003mev}. The information content of a state of a single two-level system (qubit) is given by the length of the corresponding Bloch vector, and a pure state of $N$ qubits carries $N$ bits of information. For pure states of two qubits, $I_{\mathrm{total}} = 2$. Since the information is invariant under unitary operations, we write a pure two-qubit state in its Schmidt basis
$| \psi_{ab} \rangle{=}\cos \alpha | 0_{a}0_{b} \rangle{+}\sin \alpha | 1_{a}1_{b} \rangle$. Therefore, the subsystems $a$ and $b$ are described by the same density operator and the local information reads $I_{\mathrm{local}}{=}I_a{+}I_b{=}2 \cos^2 \alpha{=}2 (\mathcal{V}_i^2{+}\mathcal{P}_i^2)$, where $i{=}a$ or $b$, and $\mathcal{V}_i$ ($\mathcal{P}_i$) denotes visibility (predictability) for the respective subsystem. Information in correlations is now given by $I_{\mathrm{corr}}{=}2{-}2 \cos^2 \alpha{=}2 \sin^2 \alpha{=}2 \mathcal{C}^2$, where $\mathcal{C}$ is the concurrence, defined by $\mathcal{C}{=}|\langle \psi | \tilde \psi \rangle|$ with $| \tilde \psi \rangle \equiv \sigma_y \otimes \sigma_y | \psi^* \rangle{=-}\cos \alpha | 11 \rangle{-}\sin \alpha | 00 \rangle$, because $| \psi^* \rangle{=}| \psi \rangle$ as the complex conjugate of $| \psi \rangle$ when the last is written in the standard basis. Putting these findings into (\ref{SOM_CI}) gives relation (2) of the main text. Figure~\ref{fig:linent}a) presents experimental verification of this traditional complementarity relation. Note that information complementarity based on linear entropy has been discussed, for example, in \cite{si:cai2007fei}.

Figure~\ref{fig:linent}b) shows that the information measure based on the linear entropy is non-additive in a sense that $I_{\mathrm{corr}}$ is negative for the product state $\rho_{\textsc{as}}$, defined in the main text. This is our reason to stick to von Neumann entropy instead of linear entropy in the main text.

\subsection{Doubly dephased states}
\begin{figure}[htbp]
\begin{center}
\includegraphics[width=0.38\textwidth]{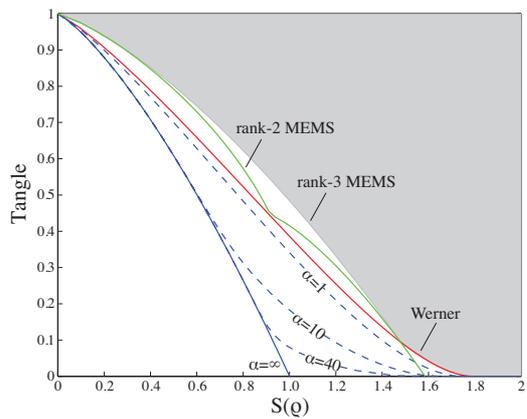}
\caption{Theory plot for states $\rho_{\textsc{d}}(\gamma)$ with $0{<}\gamma{<}1$ in comparison with Werner states $\rho_{\textsc{werner}}$ (red line) and MEMS $\rho_{\textsc{mems}}$ (green line). The blue dashed lines represent states dephased in two conjugate bases, with different weightings $\gamma$ and $\gamma^{\alpha}$. The information content of dephased states is the same as for $\rho_{\textsc{werner}}$, and larger than for the MEMS states even for entanglement arbitrarily close to zero.}
\label{fig:dephased}
\end{center}
\end{figure}

Here we present a class of weakly entangled states of two two-level systems, for which $I_{\mathrm{corr}}$ is consistently higher than that for the MEMS states. Consider the states
\begin{eqnarray}
\rho_{\textsc{d}}(\gamma) & = & \tfrac{1}{4}( \openone \otimes \openone - (1-\gamma) \sigma_x \otimes \sigma_x \\
&-& (1-\gamma) (1-\gamma^{\alpha}) \sigma_y \otimes \sigma_y - (1-\gamma^{\alpha}) \sigma_z \otimes \sigma_z), \nonumber
\end{eqnarray}
which can be seen as a result of dephasing of the Bell singlet state \ket{\psi^{-}} in the local bases of Pauli operators $\sigma_z$ and $\sigma_x$. The dephasing in the $z$-basis has strength $\gamma$, and in the $x$-basis has strength $\gamma^{\alpha}$. Just like Werner states, $\rho_{\textsc{d}}(\gamma)$ has information \emph{only} in correlations, i.e. $I_{\mathrm{local}}{=}0$, cf. Fig.~4a). We plot entanglement (tangle) of these states for various $\alpha$ in Fig.~\ref{fig:dephased}. Since the MEMS states with the same amount of total information have non-vanishing $I_{\mathrm{local}}$, they contain less information in correlations than $\rho_{\textsc{d}}(\gamma)$, even for $S(\rho){>}1$ and large $\alpha$, in which case entanglement of the dephased states is close to zero \cite{si:skerlak2009icb}. In the most extreme case, at $S(\rho){=}1$, state $\rho_{\textsc{d}}$ has zero tangle but retains $1$ bit of (classical) correlations $I_{\mathrm{corr}}$. The rank-3 MEMS (boundary of shaded area in Fig.~2), in comparison, has a tangle of $\mathcal{T}{\sim}0.48$ and $I_{\mathrm{corr}}{\sim}0.94$ bit.

\end{document}